\documentclass{emulateapj}

\slugcomment{Astronomy Letters - in press}

\shorttitle{ACCURACY OF SUPERMASSIVE BLACK HOLE MASS ESTIMATES}
\shortauthors{BOCHKAREV AND GASKELL}

\begin{document}

\title{The Accuracy of Supermassive Black Hole Masses Determined by the Single-Epoch Spectrum (Dibai) Method}

\author{NIKOLAI G. BOCHKAREV}

\affil{Sternberg Astronomical Institute, Universitetskij Prospect
13, Moscow 119991, Russia}

\author{C. MARTIN GASKELL}

\affil{Department of Astronomy, University of Texas, Austin, TX
78712-0259.}

\begin{abstract}

The first set of supermassive black hole mass estimates, published
from 1977 to 1984 by \'{E}. A. Dibai, are shown to be in excellent
agreement with recent reverberation-mapping estimates.  Comparison
of the masses of 17 AGNs covering a mass range from about  $10^6$ to
$10^9 M_{\sun}$ shows that the Dibai mass estimates agree with
reverberation-mapping mass estimates to significantly better than
$\pm 0.3$ dex and were, on average, only 0.14 dex ($\sim 40$\%)
systematically lower than masses obtained  from reverberation
mapping.  This surprising agreement with the results of over a
quarter of a century ago has important implication for the structure
and kinematics of AGNs and implies that type-1 AGNs are very
similar. Our results give strong support to the use of the
single-epoch-spectrum (Dibai) method for investigating the
co-evolution of supermassive black holes and their host galaxies.
\end{abstract}



\section{Introduction}

It is exactly 100 years this year since the publication of the first
evidence of nuclear activity in galaxies \citep{fath1908}. Over the
last half century or so, active galactic nuclei (AGNs) have been the
subject of increasingly intensive study which has resulted in tens
of thousands of papers being published. \citet{zeldovich64} and
\citet{salpeter64} proposed that the huge energy release lasting for
millions of years from a typical AGN could be explained by the
accretion of matter onto a supermassive black hole. In such
accretion, the energy output efficiency may reach 43\% of the
accreting matter's rest mass energy.  This gave the first estimates
of lower limits to the masses of AGNs because the luminosity, $L$,
cannot go much above the Eddington limit, $L_{Edd} \sim 1.3 \times
10^{38} (M/M_{\sun})$ erg/s, so the mass of the black hole in an AGN
has to be of the order of several million to several billion solar
masses, depending on the luminosity of the AGN
\citep{zeldovich+novikov64}.

While the Eddington limit gives a lower limit to the mass, $M_{BH}$,
of the black hole, the actual mass could be orders of magnitude
greater than this. To understand the working of AGNs, $M_{BH}$ needs
to be determined observationally, so estimating $M_{BH}$ has always
been considered to be a matter of utmost importance in AGN studies.
If the motions of gas clouds are dominated by gravity, masses can be
estimated in principle from the virial theorem if we know a velocity
and an appropriate distance from the center (e.g.,
\citealt{woltjer59}). The velocity along the line of sight can
easily be determined from the Doppler broadening of emission lines,
but determining the distance of the emitting material from the
central object is difficult.

It was not until the work of
\citet{dibai77,dibai78,dibai80,dibai81,dibai84a,dibai84b} that an
attempt was made at a consistent spectroscopic determination of the
masses of the central objects for a large sample of AGNs.  This
enabled Dibai to determine black hole masses and Eddington ratios,
$L/L_{Edd}$, for dozens of AGNs for the first time
\citep{dibai80,dibai84b}, and to begin investigations in what
promised to be (and indeed has turned out to be) a very fruitful
area: the relationships between these quantities and other AGN
parameters (see, for example, \citealt{dibai84a} and
\citealt{dibai+zasov85}). Unfortunately, this work was cut short by
Dibai's premature death, and in the two decades after his 1977 paper
there were only a few papers by others using the Dibai method to
estimate black hole masses (e.g.,
\citealt{joly+85,wandel+yahil85,padovani+rafanelli88}).

In the last
decade, however, there has been an enormous growth of interest in
determining masses by Dibai's method because of the close
relationship between the masses of black holes and the masses of the
bulges of their host galaxies (see \citealt{kormendy+gebhardt01} for
a review). It is only through the Dibai method that large numbers of
black hole masses in AGNs can currently be determined, and the
method has already been used in making tens of thousands of mass
estimates for AGNs of all redshifts in the Sloan Digital Sky Survey
(e.g.,
\citealt{mcclure+dunlop04,salviander+07,greene+ho07,shen+08}). It is
therefore of interest to revisit the original Dibai mass estimates
and see how they compare with more recent independent estimates.

\citet{dibai+pronik67} showed that the emitting regions of the broad
and narrow components of AGN optical emission lines (what we now
call the BLR and NLR respectively) had to originate in different
locations in space, and that the emitting gas, in both cases, had to
have a cloudy structure, that is, to fill only a small fraction,
$\epsilon$, of the volume of the corresponding region. It has long
been recognized (e.g., \citealt{bahcall+72}; see also
\citealt{bochkarev+pudenko75}) that BLR variability timescales could
be used to estimate the effective distance of the emitting gas from
the black hole, and hence to get masses via the virial theorem.

\citet{pronik+chuvaev72} presented the first long-term H$\beta$
light curve for an AGN.  \citet{lyutyi+cherepashchuk72} and
\citet{cherepashchuk+lyutyi73} made narrow-band observations of
three AGNs over several months and published the first actual
estimates of BLR sizes from the time lag between continuum
variability and line variability. \citet{bochkarev+antokhin82},
\citet{blandford+mckee82}, \citet{capriotti+82}, and
\citet{antokhin+bochkarev83} independently developed methods for
recovering information about the BLR from the response of the lines
to continuum variations, a subject which has now become known as
``reverberation mapping''. Determining BLR sizes became practical
and widespread with the introduction of cross-correlation techniques
\citep{gaskell+sparke86,gaskell+peterson87} a few years later.

Once BLR radii were being reliably obtained from cross-correlation
studies, the main remaining problem in estimating masses was in
establishing that the gas motions were dominated by gravity.  The
dominant belief from the early days of AGN studies (e.g.,
\citealt{burbidge58}) was that emission-line gas was outflowing from
AGNs, and the virial theorem obviously cannot be applied to an
outflowing wind. The idea of determining BLR kinematics diagnostics
by line-profile variations was first brought forward by S.~N.
\citet{fabrika80}, then Dibai's graduate student. The first
velocity-resolved reverberation mapping
(\citealt{gaskell88,koratkar+gaskell89}; see also
\citealt{shapovalova+01a, shapovalova+01b}) showed that the BLR was
not outflowing, but instead showing some net inflow in combination
with Keplerian and/or chaotic motion (see
\citealt{gaskell+goosmann08}). This discovery immediately permitted
the first reverberation-mapping determinations of black hole masses
\citep{gaskell88}.

Despite the promising results of the pioneering observations of
\citet{lyutyi+cherepashchuk72} and \citet{cherepashchuk+lyutyi73},
it was emphasized by \citet{bochkarev84}, confirming earlier
calculations of \citet{bochkarev+antokhin82} and
\citet{antokhin+bochkarev83}, that reliable results could only be
obtained from of long time-sequences of well-sampled, high-accuracy,
spectral and photometric observational data obtained through a
large-scale international project.  The urgent need for starting
such collaborations was discussed in detail by
\citet{bochkarev87a,bochkarev87b}.

The following decades saw the
progress of the {\it International AGN Watch} ({\it IAW}) program
(see \citealt{clavel+91,peterson+91}, et seq.) aimed at determining
the BLR size and structure from reverberation mapping. That project
proved to be one of the largest global astronomical monitoring
programs to date. More than 200 astronomers from 35 countries
cooperated for 15 years in accumulating long, densely-sampled, UV,
optical, and other wavelength time series for many AGN. As a result
of this, and of additional optical monitoring campaigns,
reverberation-mapping estimates of the AGN central black hole mass
have now been obtained for over 40 AGNs (see \citealt{peterson+04}
and \citealt{vestergaard+peterson06}).

In this paper we make a comparison of AGN central object mass values
from reverberation mapping campaigns with the masses obtained by
Dibai over a quarter of a century ago.  In Section 2 we briefly
describe the assumptions made by Dibai.  In Section 3 we compare the
results yielded by the two methods. Section 4 discusses the
implications of the comparison.

\section{The Dibai method of estimating masses}

In his first (1977) paper Dibai explained the basic principles of
his method for estimating the masses of the black holes in AGNs and
gave masses for 15 AGNs.  In 1980 and 1981 further masses for many
more AGNs were published \citep{dibai80,dibai81}. He then composed a
catalogue of the main characteristics of 77 Seyfert 1 nearby
galaxies and nearby quasars and of 24 Seyfert 2 galaxies, and used
it as the basis for new AGN mass and accretion rate estimates. What
were to be the final results were published in
\citet{dibai84a,dibai84b}, which did not appear until after the
author's untimely death occurred in November of 1983.\footnote{It
should be mentioned that significant parts of the spectral and
photometric data used by Dibai in above the mentioned papers were
obtained with Dibai's personal participation in observations: e.g.,
see \citet{dibai+81} for results of photometry for 27 AGNs and the
series of 8 papers by Arakelian, Dibai and Esipov with AGN
spectroscopy published in the journal {\it Astrophysics} in the
period 1970-1973 (see references in \citealt{dibai84b}).}

Dibai estimated the mass of the central region of AGNs under the
assumption that the gas clouds responsible for broad emission line
formation were moving with more or less parabolic velocities in the
gravitational-field of the black hole so that
\begin{equation}
M_{BH} = 1.5 R v^2/G .
\end{equation}
Here $M_{BH}$ is the mass of the central object (presumably a black
hole), $R$ is the BLR radius, $v$ is the gas velocity (which Dibai
determined from the FWHM of the broad component of H$\beta$), and
$G$ is the gravitational constant.

To estimate the size of region producing the broad component of
H$\beta$ Dibai used the simple relationship:
\begin{equation}
\epsilon (4 \pi R^3 / 3) = L_{H\beta} / E(n,T) ,
\end{equation}
where $\epsilon$ is the BLR gas filling factor; that is, the
fraction of the BLR volume filed with the emitting gas. It should be
noted that since $L \propto L_{H\beta}$ over many orders of
magnitude \citep{yee80,shuder81}, Eq(2) means that
\begin{equation}
R \propto L^{1/3}.
\end{equation}

Dibai estimated the volume occupied by the emitting gas by assuming
that the H$\beta$ line is optically thin and emitted in the
low-density approximation (the so-called ``coronal approximation'');
that is, that the results obtained for classical ionized hydrogen
zones may be applied to the line.

On the right-hand side of Eq. (2), $L(H\beta)$ is the luminosity in
the H$\beta$ line and $E(n,T) = 1.21 \times 10^{-7}$ erg cm$^{-3}$
s$^{-1}$ is emissivity for H~II zones heated up to $T = 10^4$ K with
electron number density, $n_e = 10^9$ cm$^{-3}$. Dibai based his
estimate of the density on two things.  Firstly, the absence of
broad components of forbidden spectral lines in AGN spectra leads to
a lower limit of $n_e \gtrsim 10^{8}$ cm$^{-3}$.  Secondly, the
presence of the semi-forbidden CIII] $\lambda$1909 line implies an
upper limit of $n_e \lesssim 10^{10}$ cm$^{-3}$.   For the filling
factor parameter, $\epsilon$, Dibai adopted a value obtained for the
fraction of the volume of the emitting gas in the Crab Nebula,
$\epsilon = 10^{-3}$. With these assumptions Dibai then estimated
$M_{BH}$ for almost 80 type-1 Seyfert galaxies and nearby quasars.

\begin{center}
\begin{deluxetable}{lccccc}
\tablewidth{78mm} \tablecaption{COMPARISON OF MASSES ESTIMATED BY
DIBAI WITH MASSES ESTIMATED BY REVERBERATION MAPPING} \tablehead{ &
Log ($M$) &   Log ($M$) &       &       &   }
\tablehead{\colhead{Name} & \colhead{Dibai} &   \colhead{Reverb} &
\colhead{$\pm$} & \colhead{Ref.}    & \colhead{Dibai-Reverb}}
\startdata
Mrk 335 &   7.50    &   7.15    &   0.12    &   2   &    0.35    \\
PG 0026+129 &   8.15    &   8.59    &   0.11    &   1   &   -0.44   \\
F 9 &   8.00    &   8.41    &   0.10    &   2   &   -0.41   \\
Mrk 590 &   7.40    &   7.68    &   0.07    &   2   &   -0.28   \\
3C 120  &   8.00    &   7.74    &   0.21    &   2   &    0.26    \\
Ark 120 &   8.00    &   8.18    &   0.06    &   2   &   -0.18   \\
Mrk 79  &   7.50    &   7.72    &   0.12    &   1   &   -0.22   \\
Mrk 110 &   7.20    &   7.40    &   0.11    &   2   &   -0.20   \\
NGC 3516    &   7.40    &   7.63    &   0.15    &   1   &   -0.23   \\
NGC 3783    &   7.10    &   7.47    &   0.08    &   2   &   -0.37   \\
NGC 4051    &   6.00    &   6.28    &   0.19    &   1   &   -0.28   \\
NGC 4151    &   7.20    &   7.12    &   0.16    &   2   &    0.08    \\
3C 273  &   8.50    &   8.95    &   0.09    &   2   &   -0.45   \\
Mrk 279 &   7.90    &   7.54    &   0.12    &   2   &    0.36    \\
NGC 5548    &   7.70    &   7.83    &   0.02    &   2   &   -0.13   \\
Mrk 509 &   7.70    &   8.16    &   0.04    &   2   &   -0.45   \\
NGC 7469    &   7.30    &   7.09    &   0.05    &   2   &    0.21    \\
\enddata
\tablenotetext{1}{Peterson et al.~(2004)}
\tablenotetext{2}{Vestergaard \& Peterson (2006)}
\end{deluxetable}
\end{center}


\section{Black hole masses in AGN: Dibai (1984b) vs. reverberation mapping}

Masses derived from more than 15-years of reverberation mapping of
35 AGN are given in \citet{peterson+04} and
\citet{vestergaard+peterson06}. 17 AGNs appear in both these lists
and Dibai's lists. The masses range from $M_{BH} \sim 10^6 M_{\sun}$
up to $ \sim 10^9 M_{\sun}$. The black hole mass estimates for the
AGNs in common are shown in Table 1 and plotted in Fig. 1.

\begin{figure}
\includegraphics[width=84mm]{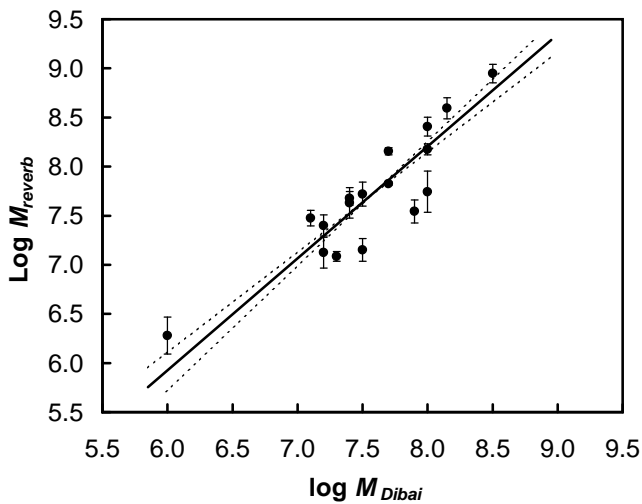} 
\caption{Comparison of the mass estimates of \citet{dibai84b},
$M_{Dibai}$ with reverberation mapping masses, $M_{reverb}$. The
solid line is the OLS-bisector fit and the two dotted lines show
regressions of (X|Y) and (Y|X).} \label{all_data}
\end{figure}

As can be seen from Fig. 1, there is obviously a good correlation.
Since there are errors in both axes we have performed an ordinary
least squares (OLS) bisector regression (see \citealt{isobe+90}).
This gives the relationship:
\begin{equation}
\log M_{reverb} = 1.14 \log M_{Dibai} - 0.85 .
\end{equation}

As can be seen from Fig. 1, there is obviously a good correlation.
Monte Carlo simulations show that the error in the slope of the
OLS-bisector line is $\pm 0.18$, so the slope does not differ
significantly from unity.   The systematic difference in the
logarithms of the masses determined by the two methods is
\begin{equation}
<\log M_{reverb} - \log M_{Dibai}>~ = -0.14 \pm 0.07 .
\end{equation}
The systematic difference between the two methods is therefore only
40\%.

The scatter in the ratio of masses determined by the two methods is
$\pm 0.28$ dex, but at least some of this must be due to the errors
in the reverberation-mapping estimates.  The mean of the formal
reverberation mass measurement errors quoted by \citet{peterson+04}
and \citet{vestergaard+peterson06} is $\pm 0.10$ dex.  If we make
the (unlikely) assumption that these measurement errors are the only
source of error in the reverberation mass estimates, then the
average error in the Dibai estimates is $\pm 0.26$ dex.  However,
this is an approximate upper limit to the errors in the Dibai method
because there are other (unknown) errors in the reverberation
mapping mass determinations.  If we assume that the errors are
equally distributed between the Dibai method and the
reverberation-mapping method, then the mean of the errors in the
Dibai method of black hole mass estimation is $\pm 0.20$ dex.
\citet{denney+09} have made a detailed examination of the effects of
random and systematic observational errors on masses estimated by
the Dibai method.  They find that with careful treatment of line
profiles using high-quality spectra (signal-to-noise ratio $\gtrsim$
20:1) it is possible to get errors of $\leqslant \pm 0.10$ dex in
the mass if the observational uncertainties are the only source of
error.  Since Dibai did not have the advantage of modern digital
spectra, the contribution of observational errors to his mass
estimates must have been significantly greater than $\pm 0.10$ dex.
Our comparison thus tells us that under ideal circumstances, the
{\em intrinsic} accuracy of the Dibai method is potentially quite
high.

\vspace{0.3cm}

\section{What does the agreement tell us?}

The surprising agreement of the estimated mass values implies that
all the ``classical'' type-1 AGNs (Seyfert 1 galaxies and nearby
quasars) are very much alike in their properties and structure, and
that object-to-object differences are smaller than has been hitherto
thought. The agreement of the Dibai and reverberation mapping
results provides support for the following:

1. BLR gas motions are dominated by gravity, as has been shown by
velocity-resolved reverberation mapping
(\citealt{gaskell88,koratkar+gaskell89}, etc.), and as is also
strongly supported by the inverse correlation of line widths with
the sizes of the emitting regions of different ions, and hence the
consistency of mass estimates from these different ions
(\citealt{krolik+91}; see also
\citealt{peterson+wandel99,peterson+wandel00}). Thus the emission
comes predominantly from gravitationally-bound gas, and not gas
flowing away from the nucleus. This is an assumption in common to
both the Dibai and reverberation mapping methods.  Dibai recognized
from the outset \citep{dibai77} that it had not yet been established
that the motions were dominated by gravity.  He also noted that
outflow velocities in Wolf-Rayet stars only slightly exceed the
escape velocity, so his mass estimates could also apply if the BLR
was a radiatively-driven outflow.

2. The kinematics of all BLRs are similar.   This is again an
assumption in common to both the Dibai and reverberation-mapping
methods.

3. The BLR size scales with optical luminosity as $R \propto
L^{\alpha}$.  This is supported by reverberation mapping estimates
of $R$ \citep{koratkar+gaskell91,peterson+wandel99,bentz+06}.  Dibai
took $\alpha$ to be $\sim 0.33$ (see Eqs. 2 and 3) while
\citet{bentz+06} get $0.52 \pm 0.04$. While the slope of the line in
Fig. 1 ($1.14 \pm 0.18$) is already consistent with unity, if we
adopt $\alpha$ = 1/2, rather than the $\alpha$ = 1/3 Dibai with
$L_{H\beta}$ (see Eqs. (2) and (3)), this changes the slope of the
line to 0.97.

4. The spectral energy distribution (SED) is very similar in all
type-1 AGNs, as has already been argued by \citet{gaskell+04} and
\citet{gaskell+benker08}.

5. The ``warm'' gas has a similar filling factor, $\epsilon \sim
0.001$ for all the type-1 AGNs. This means that the space between
BLR clouds is about 10 times bigger than the mean cloud size.  Since
the BLR is probably flattened (see \citealt{gaskell+08}), the
average separation between the clouds will be a little smaller.


6. Slow temperature variations of the ``warm'' ($T \sim 10^4$ K) BLR
gas discussed recently by \citep{popovic+08} do not make significant
deviations of the average value) of BLR gas emissivity $E(n,T)$ in
H$\beta$ line for the 17 AGNs considered here from the quantity
$1.21 \times 10^{-7}$ erg cm$^{-3}$ s$^{-1}$ adopted by Dibai
(1984b).

7. The possible presence of other BLR emission components, such as a
BLR component near the jet in radio-loud AGN (see
\citealt{bochkarev+shapovalova07,nazarova+07}, and
\citealt{arshakian+08}) in addition to the ``standard'' BLR
associated with the accretion disk does not noticeably affect the
average value of the porosity, $\epsilon$, of the gas responsible
for the BLR. The contribution of non-standard BLR components to the
total BLR emission is thus probably small, at least for the typical
BLRs of the 17 AGNs considered here.

The first two of these assumptions are common to both the Dibai
method and reverberation mapping, and these two assumptions tell us
things about BLR gas flow in AGNs.  The next two assumptions are
unique to the Dibai method and, in combination with the good
agreement of mass estimations we have found, they tell us additional
significant things about AGNs.  The third assumption is that the
physical conditions in the BLR clouds are similar (i.e., similar
densities and ionization parameters), and the fourth is that the
SEDs are similar.  The fifth assumption was only used to get the
scale factor or zero point in the original Dibai method.  The large
number of more recent applications of the Dibai method scale the
radii to reverberation mapping radii instead.  Nonetheless, the
agreement we find between Dibai's mass scale and that of
reverberation mapping implies that Dibai's assumption of a filling
factor of $\epsilon = 0.001$ is a good one. The agreement arises, of
course, because of the agreement in estimated radii.  Dibai indeed
noted at the outset \citep{dibai77} that the estimated radii agreed
with the reverberation mapping results of
\citet{lyutyi+cherepashchuk72}.

Taken together, the apparent correctness of the main assumptions of
Dibai's method reinforces the idea that, despite the wide range of
masses of their black holes and the wide range of accretion rates, a
large fraction of type-1 AGNs are surprisingly similar. This is an
important result which needs to be understood and which requires
further study.

The final, obvious, and important conclusion that can be drawn from
the good agreement we find between masses estimated by the
single-epoch spectrum method pioneered by \citet{dibai77} and by the
reverberation mapping method (and also the agreement that
\citealt{peterson+04} and \citealt{vestergaard+peterson06} find
between their own single-epoch estimate and reverberation mapping)
is that the Dibai method {\em does} give reliable black hole mass
estimates. This is very important for studying cosmic evolution of
black holes and host galaxies because it shows that the long-term,
labour-intensive monitoring of reverberation mapping is not
necessary to obtain the mass of every single AGNs in the sample. The
reliability of the Dibai method makes investigations of cosmic AGN
evolution, black hole growth, and the links with the properties and
evolution of host galaxies considerably easier. The simple
comparison we have presented here implies that we can indeed trust
this method which has now been used already for obtaining mass
estimates of many tens of thousands of AGNs.

The success of the Dibai method certainly does not mean that there
is no need for further reverberation mapping studies. Reverberation
mapping is needed for testing the reliability of black hole mass
determinations for AGNs of extreme types which are not represented
in Table 1. For example: high Eddington ratio AGNs (so called
``narrow-line Seyfert 1s'') and very low Eddington ratio AGNs (such
as FR1 radio galaxies), which might have non-classical BLRs, are not
represented. The good agreement of the Dibai method with
reverberation mapping for the 17 AGNs considered here does not
reduce the importance of the reverberation mapping in general. The
two methods are complimentary and not mutually replaceable.

\acknowledgments

This work has been supported by the Russian Foundation for Basic
Research (RFBR) through grant 06-02-16843, and by the US National
Science Foundation (NSF) through grants AST 03-07912 and AST
08-03883.  We are grateful to Misty Benz, Brad Peterson, and Xiaobo
Dong for helpful comments.


\end{document}